\documentclass[a4paper, 10pt, oneside, onecolumn, notitlepage, final]{article}
\usepackage[text={160mm, 240mm}, left=25mm, vmarginratio=1:1]{geometry}
\usepackage{graphicx}
\usepackage{amssymb, amsfonts, amsmath, amsthm, mathtools, mathrsfs}
\usepackage{upgreek}
\usepackage[usenames, dvipsnames]{xcolor}
\usepackage[colorlinks, linkcolor=blue, anchorcolor=red, citecolor=blue]{hyperref}
\usepackage[backend=biber, sorting=none, maxbibnames=3, minbibnames=3, style=numeric-comp]{biblatex}
\usepackage{braket} 
\usepackage{authblk} 
\usepackage{array}
\usepackage{bigstrut}
\usepackage{orcidlink}
\usepackage{abstract}
\usepackage{titlesec}
\usepackage[T1]{fontenc}
\usepackage[utf8]{inputenc}
\usepackage{lmodern}
\usepackage[english]{babel}

\titleformat{\section}{\large\bfseries\centering}{\thesection}{0.5 em}{}
\titleformat{\subsection}{\normalsize\bfseries\centering}{\thesubsection}{0.5 em}{}

\begin{document}

\title{\Large\bfseries Current Fluctuations in One-Dimensional Diffusion-Reaction Systems via Tensor Networks}

\author[]{\normalsize Jiayin Gu\orcidlink{0000-0002-9868-8186}\thanks{\texttt{gujiayin@njnu.edu.cn}}}
\affil[]{\normalsize School of Physics and Technology, Nanjing Normal University, Nanjing 210023, China}

\date{}
\maketitle

\begin{abstract}
Tensor networks are employed to characterize the current fluctuations in one-dimensional diffusion-reaction systems. The representative system under study is a semiconducting material where holes and electrons constitute two types of charge carriers. These holes and electrons diffuse in the system with the reactions of pair-generation and -recombination occurring between them. The system is driven by imbalanced conditions imposed at two boundaries. The large deviation function encoding the full counting statistics of electric current is numerically calculated using the density matrix renormalization group. The fluctuation theorem is shown to hold for the current. Moreover, by comparing the cases where the reactions are turned on or off, it is revealed that the reactions have a damping effect on current fluctuations. This indicates an interesting inequality, suggesting that current fluctuations are upper bounded.
\end{abstract}

\section{Introduction}

It is widely recognized that large deviation theory provides a general framework for statistical physics~\cite{Ellis_2006, Vulpiani_2014, Touchette_PhysRep_2009, Touchette_PhysicaA_2018}. For nonequilibrium systems in a steady state, the time-integrated current satisfies the large deviation principle, and all the information about its statistics is encoded in the rate function or cumulant generating function, which respectively plays the same role as entropy or free energy in equilibrium systems. Theoretical progress since the mid-1990s reveals that microreversibility leaves its footprint in the current statistics so that the associated cumulant generating function satisfies a symmetry relation, which is dubbed the fluctuation theorem~\cite{Evans_PhysRevLett_1993, Gallavotti_PhysRevLett_1996, Kurchan_JPhysA_1998, Lebowitz_JStatPhys_1999}. Starting from this theorem, Onsager reciprocal relations and their generalizations to nonlinear transport properties can be derived~\cite{Andrieux_JChemPhys_2004, Andrieux_JStatMech_2007, Gaspard_NewJPhys_2013, Barbier_JPhysA_2018}. However, more physics about the current fluctuations originating from the underlying dynamics is still hidden. In order to reveal it, comprehensive details concerning the cumulant generating function are required.

\par Large deviation theory primarily addresses the asymptotic behavior of the probabilities associated with rare events. In a nonequilibrium system at steady state, the measured average current tends to converge in probability to its expected value exponentially as the duration of the time interval approaches infinity. It is notoriously difficult to determine the rate function for this convergence, as the large deviations of the measured mean current from its expected value are exceptionally rare. Alternatively, one can compute the cumulant generating function, which is connected to the rate function by Legendre transformation. Apart from a handful of nontrivial cases where it can be calculated exactly, e.g., simple exclusion process~\cite{Mallick_JStatMech_2011, Mallick_PhysicaA_2015}, cumulant generating function can only be calculated through numerical methods. Indeed, there exist various numerical methods for computing the cumulant generating function. For stochastic processes governed by master equations, Monte Carlo methods, such as the cloning algorithm~\cite{Giardina_PhysRevLett_2006, Lecomte_JStatMech_2007, Giardina_JStatPhys_2011} and transition path sampling~\cite{Bolhuis_AnnRevPhysChem_2002}, are available. However, these simulation methods exhibit low statistical efficiency in accessing rare fluctuations, especially when applied to systems with many degrees of freedom. As is always the case, the underlying state space grows exponentially with the system size. In recent decades, tensor networks have emerged as a very promising alternative for handling such numerical complexity. The standard way to access large deviations of a dynamical observable is by computing its cumulant generating function from the leading eigenvalue of the deformed or tilted generator. In this respect, some tensor network algorithms, such as the density matrix renormalization group (DMRG)~\cite{Schollwock_RevModPhys_2005, Schollwock_AnnPhys_2011, White_PhysRevLett_1992, White_PhysRevB_1993}, are applied to first find the desired leading eigenvector represented by some tensor network states and then use them to compute the leading eigenvalue~\cite{Gorissen_PhysRevE_2009, Gorissen_PhysRevE_2012, Gorissen_PhysRevLett_2012, Banuls_PhysRevLett_2019, Helms_PhysRevE_2019, Helms_PhysRevLett_2020, Gu_NewJPhys_2022, Strand_JChemPhys_2022a, Strand_JChemPhys_2022b}. Furthermore, tensor networks also find applications in nonequilibrium physics in many other respects. For example, they have been used to calculate work statistics for quantum many-body systems under prescribed driven protocols~\cite{Gu_PhysRevRes_2022, Lin_PhysRevRes_2024}, to study the heat transfer in non-Markovian open quantum systems~\cite{Popovic_PRXQuantum_2021}, or to efficiently sample rare events~\cite{Causer_PhysRevE_2021, Causer_PhysRevLett_2023}. In particular, for the stochastic dynamics in diffusion-reaction systems, they have been used to investigate the nonequilibrium critical phenomena~\cite{Carlon_EurPhysJB_1999} or to compute the rate of switching between metastable macrostates in stochastic diffusion-reaction dynamics~\cite{Nicholson_PhysRevX_2023}.

\par In this work, we extend the application of tensor networks to calculate the cumulant generating function of current statistics in a one-dimensional (1D) diffusion-reaction system. In previous works on diffusion-reaction systems investigated with tensor networks~\cite{Carlon_EurPhysJB_1999, Nicholson_PhysRevX_2023}, there exists only one reactant species with an unfixed concentration. However, in the system studied in this work, two types of particles are present. They undergo diffusion with reactions constantly occurring between them. As a consequence, their concentrations fluctuate, and two stochastic variables are needed to characterize the local state of the system. This peculiarity poses challenge in the application of tensor networks. In this work, we use composite indices to label the system's local state. This constitutes one of the novel aspects of this work. The price to pay is that each composite index needs to take more discrete values, which demands more computational cost. In practical tensor-network calculations, this is acceptable; new physics can still be revealed in the system. The fluctuating current across the system is induced by imbalanced boundary conditions. In this work, we focus on how the current fluctuations are influenced by the reactions. After the procedure of spatial discretization, a master equation is established to describe the stochastic evolution of the system state and, thereby, the full counting statistics for the current is performed. The distribution function of the system state is represented by a matrix product state (MPS) and its tilted generator by a matrix product operator (MPO). The DMRG algorithm is employed to compute the leading eigenvalue of the tilted generator, giving the cumulant generating function.

\begin{figure}
\centering
\begin{minipage}[t]{0.6\hsize}
\resizebox{1.0\hsize}{!}{\includegraphics{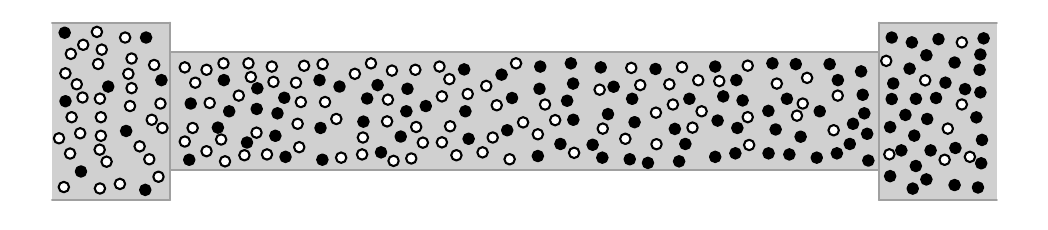}}
\end{minipage}
\caption{Schematic diagram of a 1D semiconducting material. The white dots represent holes and the black ones represent electrons. The system is in contact at the ends with two reservoirs that fix the densities of holes and electrons.}
\label{fig_system}
\end{figure}

\par This paper is organized as follows. In Sec.~\ref{sec_semiconductor}, we introduce a stochastic description of the physical processes taking place inside a 1D diffusion-reaction system. In Sec.~\ref{sec_TN}, we turn to the tensor-network formulation of the stochastic description. In particular, a significant effort is devoted to the construction of local operators describing elementary jump processes. The results are summarized in Sec.~\ref{sec_results}, where current fluctuations are shown to be upper bounded through the detailed analysis of the cumulant generating function. Conclusion and perspectives are given in Sec.~\ref{sec_conclusion}. In addition, some technical details on how to perform tensor-network calculations are presented in two Appendixes.

\section{One-Dimensional Semiconducting Material}\label{sec_semiconductor}

The system we consider is a typical diffusion-reaction system of physical interest: a semiconducting material, as shown in Fig.~\ref{fig_system}. The system is a three-dimensional rod of length $l$ extending along $x$ axis from $0$ to $l$ and with a section area $\Sigma$ in the transverse $y$ and $z$ directions. The mobile charge carriers distributed inside the system are positively charged holes and negatively charged electrons with their static densities expressed as functions of position $p(x)$ and $n(x)$. The two terminals of the system are in contact with the reservoirs containing fixed concentrations of holes and electrons, $\bar{p}_{\rm L}$, $\bar{p}_{\rm R}$, $\bar{n}_{\rm L}$, and $\bar{n}_{\rm R}$. The diffusion of the charge carriers is quantified by their respective coefficients that are assumed equal here, $D_p=D_n$. Moreover, holes and electrons are generated and recombined through the reaction
\begin{align}
\emptyset\xrightleftharpoons[k_-]{k_+} h^++e^- \text{,}
\end{align}
where $k_+$ and $k_-$ are respectively the hole-electron pair generation and recombination rate constants. For simplicity, we neglect the long-range electrostatic interactions between charge carriers, only taking into account the physical processes of diffusion and reaction.

\subsection{Stochastic Diffusion-Reaction Equations}

The thermal agitation inside the system generates incessant erratic movements of holes and electrons, in turn causing local fluctuations in hole diffusion, electron diffusion, and reactions. Above the nanoscale, the diffusion and reaction can be described in terms of stochastic processes~\cite{Nicolis_1977, Andrieux_JStatMech_2009, Gaspard_JChemPhys_2004}. The balance equations for holes and electrons, with respective densities $p(x,t)$ and $n(x,t)$, can be expressed as
\begin{align} 
& \frac{\partial p(x,t)}{\partial t}+\frac{\partial j_p(x,t)}{\partial x}=\sigma(x,t) \text{,} \label{eq_sto1} \\
& \frac{\partial n(x,t)}{\partial t}+\frac{\partial j_n(x,t)}{\partial x}=\sigma(x,t) \text{,} \label{eq_sto2}
\end{align}
where the current densities are given by
\begin{align}
& j_p(x,t)=-D_p\frac{\partial p(x,t)}{\partial x}+\delta j_p(x,t) \text{,} \label{eq_sto3} \\
& j_n(x,t)=-D_n\frac{\partial n(x,t)}{\partial x}+\delta j_n(x,t) \text{,} \label{eq_sto4}
\end{align}
and the reaction rate density by
\begin{align}
\sigma(x,t)=k_+-k_-p(x,t)n(x,t)+\delta\sigma(x,t) \text{.}
\end{align}
Here, the terms $\delta j_p(x,t)$, $\delta j_n(x,t)$ and $\delta\sigma(x,t)$ are Gaussian white noise fields characterized by
\begin{align}
& \langle \delta j_p(x,t)\rangle=\langle\delta j_n(x,t)\rangle=\langle\delta\sigma(x,t)\rangle=0 \text{,} \\
& \langle \delta j_p(x,t)\otimes\delta j_p(x',t')\rangle\!=\!\Gamma_p(x,t)\delta(x-x')\delta(t-t') \text{,} \\
& \langle \delta{j}_n(x,t)\otimes\delta j_n(x',t')\rangle\!=\!\Gamma_{n}(x,t)\delta(x-x')\delta(t-t') \text{,} \\
& \langle\delta\sigma(x,t)\otimes\delta\sigma(x',t')\rangle=\Gamma_{\sigma}(x,t)\delta(x-x')\delta(t-t') \text{,} \\
& \langle \delta j_p(x,t)\otimes\delta j_n(x',t')\rangle=0 \text{,} \\
& \langle\delta\sigma(x,t)\otimes\delta j_n(x',t')\rangle\!=\!\langle\delta\sigma(x,t)\otimes\delta j_p(x',t')\rangle\!=\!0 \text{,}
\end{align}
where $\langle\cdot\rangle$ denotes the statistical average, and the spectral densities for hole diffusion, electron diffusion, and reaction are given by
\begin{align}
& \Gamma_p(x,t)\equiv 2D_pp(x,t) \text{,} \\
& \Gamma_n(x,t)\equiv 2D_nn(x,t) \text{,} \\
& \Gamma_{\sigma}(x,t)\equiv k_++k_-p(x,t)n(x,t) \text{.}
\end{align}
In the reservoirs, the hole-electron pair generation should be balanced with the recombination, so the condition
\begin{align}  
k_+=k_-\bar{p}_{\rm L}\bar{n}_{\rm L}=k_-\bar{p}_{\rm R}\bar{n}_{\rm R}
\end{align}
should be satisfied. In Eqs.~(\ref{eq_sto3}) and (\ref{eq_sto4}), we have neglected the contribution of drift induced by the electric field, retaining only the terms of diffusion. The full description of the above diffusion-reaction system, including electrostatic interactions, has been explored in the extensive studies on diodes and transistors~\cite{Gu_PhysRevE_2018, Gu_PhysRevE_2019, Gu_PhysRevE_2025a}. The Eqs.~(\ref{eq_sto1})-(\ref{eq_sto4}) constitute a set of stochastic diffusion-reaction equations. The advantage of this approach is that the usual phenomenological parameters suffice for the stochastic description.

\par By averaging the noise and discarding the time derivative terms, the mean-field equations under stationary conditions can therefore be obtained:
\begin{align}
& \frac{{\rm d}p(x)}{{\rm d}x}=-\frac{j_p(x)}{D_p} \text{,} \label{eq_ode1} \\
& \frac{{\rm d}n(x)}{{\rm d}x}=-\frac{j_n(x)}{D_n} \text{,} \label{eq_ode2} \\
& \frac{{\rm d}j_p(x)}{{\rm d}x}=k_+-k_-p(x)n(x) \text{,} \label{eq_ode3} \\
& \frac{{\rm d}j_n(x)}{{\rm d}x}=k_+-k_-p(x)n(x) \text{,} \label{eq_ode4}
\end{align}
which are a set of ordinary differential equations (ODEs) subject to the boundary conditions $p(0)=\bar{p}_{\rm L}$, $p(l)=\bar{p}_{\rm R}$, $n(0)=\bar{n}_{\rm L}$, and $n(l)=\bar{n}_{\rm R}$. For simplicity, we impose constraints:
\begin{align}
\bar{p}_{\rm L}=\bar{n}_{\rm R} \text{,}\hspace{1cm} \bar{p}_{\rm R}=\bar{n}_{\rm L} \text{,} \label{eq_constraints}
\end{align}
so that the boundary conditions are symmetric under inversion and permutation between holes and electrons. The detailed explanation for this will be provided in the following text. The terms $p(x)n(x)$ appearing in Eqs.~(\ref{eq_ode3}) and (\ref{eq_ode4}) render the ODEs nonlinear. As such, numerical routines are necessary to solve them.

\subsection{Master Equations}

\par The time evolution of the distribution of holes and electrons in the system can also be described as a Markov jump process, which is formulated in terms of a master equation. This is more amenable to numerical simulation. For this purpose, we spatially discretize the system into $L$ cells of width $\Delta x$ and volume $\Omega$. Thus, the mesoscopic state of the system is specified by the hole numbers ${\bf P}=\{P_i\}_{i=1}^L$ and electron numbers ${\bf N}=\{ N_i\}_{i=1}^L$ in these cells. The left (respectively, right) reservoir is modeled as a cell containing fixed numbers of holes and electrons $\bar{P}_{\rm L}\equiv P_0=\bar{p}_{\rm L}\Omega$, $\bar{N}_{\rm L}\equiv N_0=\bar{n}_{\rm L}\Omega$ (respectively, $\bar{P}_{\rm R}\equiv P_{L+1}=\bar{p}_{\rm R}\Omega$, $\bar{N}_{\rm R}\equiv N_{L+1}=\bar{n}_{\rm R}\Omega$). Here, for notational consistency, the two reservoir cells are also indexed with $i=0$ and $i=L+1$, respectively. In this discretization scheme, the system randomly jumps between its states according to the network
\begin{equation}
\resizebox{0.94\hsize}{!}{$
\begin{array}{ccccccccccccc}
\bar{P}_{\rm L} & \xrightleftharpoons[W_0^{(-P)}]{W_0^{(+P)}} & P_1 & \xrightleftharpoons[W_1^{(-P)}]{W_1^{(+P)}} & P_2 & \xrightleftharpoons[W_2^{(-P)}]{W_2^{(+P)}} & \cdots & \xrightleftharpoons[W_{L-2}^{(-P)}]{W_{L-2}^{(+P)}} & P_{L-1} & \xrightleftharpoons[W_{L-1}^{(-P)}]{W_{L-1}^{(+P)}} & P_L & \xrightleftharpoons[W_L^{(-P)}]{W_L^{(+P)}} & \bar{P}_{\rm R} \\
 & & {\scriptstyle W_1^{(+)}}\updownarrow{\scriptstyle W_1^{(-)}} & & {\scriptstyle W_2^{(+)}}\updownarrow{\scriptstyle W_2^{(-)}} & & \cdots & &{\scriptstyle W_{L-1}^{(+)}}\updownarrow{\scriptstyle W_{L-1}^{(-)}} & & {\scriptstyle W_L^{(+)}}\updownarrow{\scriptstyle W_L^{(-)}} & &  \\
\bar{N}_{\rm L} & \xrightleftharpoons[W_0^{(-N)}]{W_0^{(+N)}} & N_1 & \xrightleftharpoons[W_1^{(-N)}]{W_1^{(+N)}} & N_2 & \xrightleftharpoons[W_2^{(-N)}]{W_2^{(+N)}} & \cdots & \xrightleftharpoons[W_{L-2}^{(-N)}]{W_{L-2}^{(+N)}} & N_{L-1}& \xrightleftharpoons[W_{L-1}^{(-N)}]{W_{L-1}^{(+N)}} & N_L & \xrightleftharpoons[W_L^{(-N)}]{W_L^{(+N)}} & \bar{N}_{\rm R} 
\end{array}$} \text{.} \label{eq_network}
\end{equation}
The probability ${\cal P}({\bf P},{\bf N},t)$ that the cells contain the particle numbers ${\bf P}=\{P_i\}_{i=1}^L$ and ${\bf N}=\{N_i\}_{i=1}^L$ at time $t$ is ruled by the master equation
\begin{align}
\frac{{\rm d}{\cal P}}{{\rm d}t}=\hat{L}{\cal P}=& \sum_{i=0}^{L} \Biggl[\left({\rm e}^{+\partial_{P_i}}{\rm e}^{-\partial_{P_{i+1}}}-1\right)W_i^{(+P)}{\cal P}+\left({\rm e}^{-\partial_{P_i}}{\rm e}^{+\partial_{P_{i+1}}}-1\right)W_i^{(-P)}{\cal P} \nonumber \\
& \qquad +\left({\rm e}^{+\partial_{N_i}}{\rm e}^{-\partial_{N_{i+1}}}-1\right)W_i^{(+N)}{\cal P}+\left({\rm e}^{-\partial_{N_i}}{\rm e}^{+\partial_{N_{i+1}}}-1\right)W_i^{(-N)}{\cal P}\Biggr] \nonumber \\
& +\sum_{i=1}^{L}\Biggl[\left({\rm e}^{-\partial_{P_i}}{\rm e}^{-\partial_{N_i}}-1\right)W_i^{(+)}{\cal P}+\left({\rm e}^{+\partial_{P_i}}{\rm e}^{+\partial_{N_i}}-1\right)W_i^{(-)}{\cal P}\Biggr] \text{,} \label{eq_master_equation}
\end{align}
where $\hat{L}$ denotes the generator driving the Markov jump process, and ${\rm e}^{\pm\partial_X}$ (with $X=P,N$) are raising and lowering operators such that for any function $f(X)$ we have ${\rm e}^{\pm\partial_X}f(X)=f(X\pm 1)$. This relation can be established by expanding both sides into Taylor series. The transition rates for charge carriers hopping between neighboring cells are given by
\begin{align}
& W_i^{(+P)}=kP_i \text{,} & W_i^{(-P)}=kP_{i+1} \text{,} \\
& W_i^{(+N)}=kN_i \text{,} & W_i^{(-N)}=kN_{i+1} \text{,}
\end{align}
which are linearly dependent on the charge carrier numbers in the departure cell. The common prefactor $k\equiv D_p/\Delta x^2=D_n/\Delta x^2$ guarantees the consistency with the macroscopic description of the diffusion process. For the reaction process, the transition rates are
\begin{align}
& W_i^{(+)}=k_+\Omega \text{,} & W_i^{(-)}=k_-N_iP_i/\Omega \text{,}
\end{align}
where we see the nonlinearity in the charge carrier numbers in the pair-recombination rate. This makes the master equation~(\ref{eq_master_equation}) analytically intractable. Because the hole and electron numbers in reservoir cells are maintained in constant, the raising and lowering operators acting on the reservoir cells must be ${\rm e}^{\pm\partial_{P_0}}={\rm e}^{\pm\partial_{N_0}}={\rm e}^{\pm\partial_{P_{L+1}}}={\rm e}^{\pm\partial_{N_{L+1}}}=1$. When the actions of these raising and lowering operators are explicitly expressed, the master equation~(\ref{eq_master_equation}) transforms into a more classical form, which is however less compact.

\par The master equation~(\ref{eq_master_equation}) can be simulated using the Gillespie algorithm~\cite{Gillespie_JComputPhys_1976} for generating random trajectories. In the limit $P_i\gg1$ and $N_i\gg 1$, the master equation~(\ref{eq_master_equation}) can be approximately reduced to the Fokker-Planck equation through the Kramers-Moyal expansion. This further enables a much faster simulation based on the stochastic process of Langevin type. Such a numerical route has been used to simulate the stochastic process of charge carriers in diodes and transistors~\cite{Gu_PhysRevE_2018, Gu_PhysRevE_2019, Gu_PhysRevE_2025a}. In these studies, both the fundamental issue of microreversibility and the functionalities of these electronic devices are primarily focused on. The fluctuation relations implied by the microreversibility were tested directly in near-equilibrium regimes or indirectly with the a developed coarse-grained model~\cite{Gu_JStatMech_2020} in medium-from-equilibrium regimes. Since these studies were based on numerical simulations at the trajectory level, the cumulant generating function containing full information about the current statistics has never been obtained. In this work, we switch to a different route, where the master equation~(\ref{eq_master_equation}) is integrated at the ensemble level of probability distribution. However, the problem encountered here is that the state space is exponentially large. Suppose that the number of holes or electrons in each cell is allowed to take discrete values from $0$ to $M-1$ after truncation by imposing a cutoff threshold, the whole composite system would have a space of $M^{2L}$ states. This is not numerically tolerable if we integrate the master equation~(\ref{eq_master_equation}) with matrices and/or vectors. We will see in Sec.~\ref{sec_TN} that tensor networks are the tools that circumvent this difficulty and, consequently, the cumulant generating function for current statistics can be successfully evaluated.

\begin{figure}
\centering
\begin{minipage}[t]{0.49\hsize}
\resizebox{1.0\hsize}{!}{\includegraphics{MPS.pdf}}
\end{minipage}
\begin{minipage}[t]{0.49\hsize}
\resizebox{1.0\hsize}{!}{\includegraphics{MPO.pdf}}
\end{minipage}
\caption{Graphical notations of the distribution function $F_{\bf PN}$ represented as an MPS (left panel) and the tilted generator $\hat{L}_{\lambda{\bf P'N'}}^{\bf PN}$ represented as an MPO (right panel). The composite indices $\{P_iN_i\}$ are used to describe the state of each discretized cell. Only five sites are shown for illustration. In the MPO, the counting parameter $\lambda$ is included in the first site.}
\label{fig_TN}
\end{figure}

\subsection{Current Fluctuations}

Under the imbalanced boundary conditions, the experimentally measurable quantity is the electric current contributed by both holes and electrons. Here we consider the fluctuating electric current crossing the section between the left reservoir and the system. The instantaneous electric current can be expressed as
\begin{align}
{\cal I}(t)=\sum_{n=-\infty}^{+\infty}q_n\delta(t-t_n) \text{,}
\end{align}
with $q_n=\pm e$ depending on whether the charge carrier is hole or electron and moves into or out of the system. Here $e=|e|$ is the elementary charge and $\{t_n\}$ are the time instants when these charge carriers cross the section. The number of cumulative carrier transfers over the time interval $[0,\,t]$ is defined as
\begin{align}
Z(t)=\frac{1}{e}\int_0^t{\cal I}(t'){\rm d}t' \text{,}
\end{align}
which is a fluctuating quantity. The probability distribution of $Z$ measured over such a time interval is expected to satisfy the fluctuation relation~\cite{Andrieux_JStatPhys_2007, Andrieux_JStatMech_2009}
\begin{align}
\frac{{\cal P}(Z,t)}{{\cal P}(-Z,t)}\simeq_{t\to\infty}\exp\left(AZ\right) \text{,} \label{eq_FT1}
\end{align}
where $A$ is the affinity driving the system out of equilibrium~\cite{DeDonder_1936}. It is given by
\begin{align}
A=\ln\frac{\bar{p}_{\rm L}}{\bar{p}_{\rm R}}=\ln\frac{\bar{n}_{\rm R}}{\bar{n}_{\rm L}} \text{,} \label{eq_A}
\end{align}
which can be evaluated with the Schnakenberg graph analysis~\cite{Schnakenberg_RevModPhys_1976}. The requirement for consistency of the affinity in terms of both hole and electron distributions motivates the constraints~(\ref{eq_constraints}) on the boundary conditions. The cumulant generating function in terms of the counting parameter $\lambda$ can be defined,
\begin{align}
Q(\lambda)\equiv\lim_{t\to\infty}-\frac{1}{t}\ln\sum_Z{\cal P}(Z,t)\,{\rm e}^{-\lambda Z} \text{,} \label{eq_Q}
\end{align}
which is the large-deviation function containing full information on current fluctuations. According to the fluctuation relation~(\ref{eq_FT1}), the cumulant generating function satisfies the relation
\begin{align}
Q(\lambda)=Q(A-\lambda) \text{,} \label{eq_FT2}
\end{align}
which is called Gallavotti-Cohen symmetry~\cite{Evans_PhysRevLett_1993, Gallavotti_JStatPhys_1995, Lebowitz_JStatPhys_1999}. The mean current and its diffusivity can be obtained by taking successive derivatives with respect to the counting parameter,
\begin{align}
& J\equiv\lim_{t\to\infty}\frac{1}{t}\langle Z(t)\rangle=\left.\frac{\partial Q(\lambda)}{\partial\lambda}\right\vert_{\lambda=0} \text{,} \label{eq_J} \\
& D\equiv\lim_{t\to\infty}\frac{1}{2t}\langle(Z(t)-Jt)^2\rangle=\left.-\frac{1}{2}\frac{\partial^2Q(\lambda)}{\partial\lambda^2}\right\vert_{\lambda=0} \text{.} \label{eq_D}
\end{align}
Accurate determination of the mean current, as well as the higher-order cumulants, is crucial for unveiling the nonequilibrium properties originating from the underlying stochastic dynamics. In this work, we are primarily interested in how reactions between holes and electrons influence the total electric current and its fluctuations.

\par According to Eqs.~(\ref{eq_J}) and (\ref{eq_D}), the mean current and its diffusivity can be statistically evaluated according to their definitions from a sample of stochastic trajectories. This method of obtaining the accurate the mean current and its diffusivity heavily relies on sufficiently large samples. Instead, we can first calculate the cumulant generating function and then evaluate the cumulants from successive derivatives with respect to the counting parameter. The method for calculating the cumulant generating function is called the full counting statistics. For this purpose, we can include the counting parameter in the generator to obtain the tilted generator $\hat{L}_{\lambda}$. Full counting statistics essentially boils down to calculating the dominant eigenvalue of $\hat{L}_{\lambda}$. We will see in the following sections that this way of evaluating the mean current and its diffusivity is advantageous, especially when tensor networks are employed.

\section{Tensor-Network Formulation}\label{sec_TN}

\par Tensor-network techniques are state-of-the-art numerical methods for studying strongly correlated, quantum many-body systems~\cite{Ran_2020, Banuls_AnnuRevCondensMatterPhys_2023, Orus_NatRevPhys_2019}. They are capable of investigating quantum systems that go beyond mean-field approximations and access large system sizes. Their success is largely due to the fact that the quantum states of physical relevance have low bipartite entanglement that satisfies the area law~\cite{Eisert_RevModPhys_2010}. The most celebrated tensor-network technique is DMRG~\cite{White_PhysRevLett_1992, White_PhysRevB_1993, Schollwock_AnnPhys_2011}, which allows for high-precision calculations of ground state and low-energy spectrum of a quantum many-body Hamiltonian. Other tensor-network techniques, such as the time evolving block decimation (TEBD)~\cite{Vidal_PhysRevLett_2003, Vidal_PhysRevLett_2004} and the time dependent variational principle (TDVP)~\cite{Haegeman_PhysRevLett_2011, Haegeman_PhysRevB_2016} can also be used to obtain the ground state by imaginary time evolution. In TEBD algorithm, however, two neighboring sites are updated simultaneously. This can be numerically expensive if the local degrees of freedom are very high. The TDVP algorithm is quite similar to DMRG~\cite{Haegeman_PhysRevB_2016}. Both can be implemented to update a single site at a time. For the purpose of the present work, TDVP and DMRG can be used to achieve the same goal. However, DMRG is more popular and is also a little bit simpler to implement numerically. In this work, the latter is used in practice.

\par Similarities can be drawn between quantum many-body systems and the classical many-body system discussed in this work. In both scenarios, the underlying state space grows exponentially with the degrees of freedom. The tensor-network techniques to compute the ground state energy for quantum systems are apparently an eigenvalue problem. The computation of cumulant generating function from the leading eigenvalue of tilted generator is in spirit very much the same. This motivates the use of tensor-network techniques for the calculation of cumulant generating function. However, we are also aware of two key differences: (1) the probability distributions for classical systems normalize in a different way from the quantum wave functions and (2) the tilted generator is inherently nonsymmetric or non-Hermitian while quantum Hamiltonian is always Hermitian. In the following, we formulate the master equation~(\ref{eq_master_equation}) in terms of tensor networks and utilize the DMRG algorithm, which is tailored to handle these differences.

\subsection{Matrix Product States}

The core idea of DMRG is to represent the quantum state of a 1D many-body system using a variational ansatz called MPS. This also applies to the classical state of 1D many-body systems undergoing Markov jump process. For this purpose, we first introduce notation $\ket{P_iN_i}$ or its adjoint $\bra{P_iN_i}$ with composite index $P_iN_i$ for the local state where there are $P_i$ holes and $N_i$ electrons in the $i$-th cell. Similarly, the system state is denoted by $\ket{\bf PN}\equiv\ket{P_iN_i}^{\otimes_{i=1}^L}$ or $\bra{\bf PN}\equiv\bra{P_iN_i}^{\otimes_{i=1}^L}$. Following the quantum-mechanical convention, the normalization and orthogonality are assumed,
\begin{align}
& \braket{P_i'N_i'|P_iN_i}=\delta_{P_i',P_i}\delta_{N_i',N_i} \text{,} \\
& \braket{{\bf P}'{\bf N}'|{\bf PN}}=\delta_{{\bf P}',{\bf P}}\delta_{{\bf N}',{\bf N}} \text{.} 
\end{align}
In this way, an arbitrary distribution function $F({\bf PN})$ can be expressed as
\begin{align}
\ket{\bf F}=\sum_{\bf PN}F({\bf PN})\ket{\bf PN} \text{,}
\end{align}
and the normalization is also required,
\begin{align}
\braket{{\bf F}^{\dagger}|{\bf F}}=\sum_{\bf PN}F^{\dagger}({\bf PN})F({\bf PN})=1 \text{.} \label{eq_normalization}
\end{align}
Since the distribution function $F({\bf PN})$ is real in this context, we actually have $F^{\dagger}({\bf PN})=F({\bf PN})$. The distribution function can be represented as an MPS~\cite{Verstraete_AdvPhys_2008, PerezGarcia_QuantumInfComput_2007, Schollwock_AnnPhys_2011},
\begin{align}
\ket{{\bf F}}=\sum_{\{P_iN_i\}=1}^{M^2}{\rm Tr}\left(F_1^{P_1N_1}\cdots F_L^{P_LN_L}\right)\ket{{\bf PN}} \text{,}
\end{align}
where each $F_i^{P_iN_i}$ is a $d_{i-1}\times d_i$ matrix. Here $\{d_i\}$ are bond dimensions quantifying the state correlations between neighboring cells. The graphical representation of an MPS is shown in the left panel of Fig.~\ref{fig_TN}. In the DMRG calculation, if the finally obtained distribution function $F({\bf PN})$ represents the probability distribution, then the probability distribution can be constructed as
\begin{align}
{\cal P}({\bf PN})=\frac{F({\bf PN})}{\sum_{\bf PN}F({\bf PN})} \text{,} \label{eq_normalization2}
\end{align}
so that the probability normalization $\sum_{\bf PN}{\cal P}({\bf PN})=1$ can be guaranteed.

\subsection{Matrix Product Operators}
 
\par The most crucial step in implementing DMRG calculations is to represent the tilted generator $\hat{L}_{\lambda}$ as an MPO~\cite{Chan_JChemPhys_2016}. For this purpose, we now turn to the Doi-Peliti formalism~\cite{Doi_JPhysAMathGen_1976, Peliti_JPhys_1985}, which is a classical version of the second-quantization methods from quantum field theory. In this framework, we introduce local operators accounting for elementary jump processes in the network~(\ref{eq_network}). The particle transitions between two neighboring cells can be associated with a local annihilation operator acting on one cell and a local creation operator acting the other. According to this reasoning, we define the annihilation operator for holes $a_i^-$, creation operator for holes $a_i^+$, annihilation operator for electrons $b_i^-$, and creation operator for electrons $b_i^+$, by
\begin{align}
& \braket{P_i'N_i'|a_i^-|P_iN_i}=kP_i\delta_{P_i',P_i-1}\delta_{N_i',N_i} \text{,} \label{eq_a_minus} \\
& \braket{P_i'N_i'|a_i^+|P_iN_i}=\delta_{P_i',P_i+1}\delta_{N_i',N_i} \text{,} \label{eq_a_plus} \\
& \braket{P_i'N_i'|b_i^-|P_iN_i}=kN_i\delta_{P_i',P_i}\delta_{N_i',N_i-1} \text{,} \label{eq_b_minus} \\
& \braket{P_i'N_i'|b_i^+|P_iN_i}=\delta_{P_i',P_i}\delta_{N_i',N_i+1} \text{,} \label{eq_b_plus}
\end{align}
where $kP_i$ and $kN_i$ appearing in the definitions of local annihilation operators manifest their meaning of transition rates. Furthermore, we define local operators accounting for the probability loss, $a_i$ for holes, and $b_i$ for electrons, by
\begin{align}
\braket{P_i'N_i'|a_i|P_iN_i}=kP_i\delta_{P_i',P_i}\delta_{N_i',N_i} \text{,} \label{eq_a} \\
\braket{P_i'N_i'|b_i|P_iN_i}=kN_i\delta_{P_i',P_i}\delta_{N_i',N_i} \text{,} \label{eq_b}
\end{align}
which are here intuitively dubbed as particle number operators. It should be pointed out that the operator definitions by Eqs.~(\ref{eq_a_minus})-(\ref{eq_b_plus}) and Eqs.~(\ref{eq_a}) and (\ref{eq_b}) are only applicable to the cells for the intermediate system, $1\le i\le L$. For the reservoir cells at the boundaries, $i=0,L+1$, these operators should be defined separately,
\begin{align}
& a_0^-=a_0=k\bar{P}_{\rm L} \text{,} \label{eq_bc1} \\
& a_{L+1}^-=a_{L+1}=k\bar{P}_{\rm R} \text{,} \label{eq_bc2} \\
& b_0^-=b_0=k\bar{N}_{\rm L}  \text{,} \label{eq_bc3} \\
& b_{L+1}^-=b_{L+1}=k\bar{N}_{\rm R} \text{,} \label{eq_bc4} \\
& a_0^+=b_0^+=a_{L+1}^+=b_{L+1}^+=1 \text{,} \label{eq_bc5}
\end{align}
which are all constants. The reason for this is obviously that the particle numbers in the reservoir cells are maintained constant. Similarly, for the jumps associated with reactive events in each cell, $1\le i\le L$, we may also define the so-called recombination and generation operators, as well as the associated operator for probability loss, $c_i^-$, $c_i^+$, $c_i$ by
\begin{align}
& \braket{P_i'N_i'|c_i^-|P_iN_i}=k_-\frac{P_iN_i}{\Omega}\delta_{P_i',P_i-1}\delta_{N_i',N_i-1} \text{,} \label{eq_c_minus} \\
& \braket{P_i'N_i'|c_i^+|P_iN_i}=k_+\Omega\delta_{P_i',P_i+1}\delta_{N_i',N_i+1} \text{,} \label{eq_c_plus} \\
& \braket{P_i'N_i'|c_i|P_iN_i}=\left(k_+\Omega+k_-\frac{P_iN_i}{\Omega}\right)\delta_{P_i',P_i}\delta_{N_i',N_i} \text{.} \label{eq_c}
\end{align}
At this stage, all necessary local operators have been defined. Because the number of charge carriers in a single cell is unlimited in theory, the matrix representation of these local operators has infinite dimensions. Thus, in following numerical calculations, truncation of the matrix is necessary.

The generator in the master equation~(\ref{eq_master_equation}) including the counting parameter $\lambda$ can now be expressed independently as follows,
\begin{align}
\hat{L}_{\lambda}= & a_0^-\otimes a_1^+{\rm e}^{-\lambda}+a_0^+\otimes a_1^-{\rm e}^{+\lambda}+\sum_{i=1}^L\left(a_i^-\otimes a_{i+1}^++a_i^+\otimes a_{i+1}^-\right)-\sum_{i=0}^L\left(a_i+a_{i+1}\right) \nonumber \\
& + b_0^-\otimes b_1^+{\rm e}^{+\lambda}+b_0^+\otimes b_1^-{\rm e}^{-\lambda}+\sum_{i=1}^L\left(b_i^-\otimes b_{i+1}^++b_i^+\otimes b_{i+1}^-\right)-\sum_{i=0}^L\left(b_i+b_{i+1}\right) \nonumber \\
& + \sum_{i=1}^L\left( c_i^{+}+c_i^{-}-c_i\right) \text{,} \label{eq_tilted_generator}
\end{align}
which bears a form similar to the second-quantization of many-body Hamiltonians. The parameter $\lambda$ included in Eq.~(\ref{eq_tilted_generator}) counts the charge transfers crossing the section between the left reservoir and the intermediate system. Since the current is composed of holes and electrons, a proper counting scheme is assumed depending on whether the carrier is positively or negatively charged and jumps in or out of the system. The tilted generator $\hat{L}_{\lambda}$ expressed in Eq.~(\ref{eq_tilted_generator}) can now be easily represented as an MPO, as schematically shown in right panel of Fig.~\ref{fig_TN}. This can be automated with the ITensors package~\cite{Fishman_SciPostPhysCodeb_2022} in the \textsc{julia} programming language~\cite{Bezanson_SIAMRev_2017}. A more detailed account of how to construct MPO can be found in Appendix~\ref{app_MPSMPO}.

\begin{table}
\caption{The parameter values adopted in DMRG calculations.}
\begin{center}
\begin{tabular}{>{\centering\arraybackslash}m{3.5cm}>{\centering\arraybackslash}m{3.5cm}}
\hline
\hline
$D_p=D_n=0.01$   &  $\Delta x=0.1$  \bigstrut \\ \hline
$\Omega=4$  &  $L=10$  \bigstrut \\ \hline
$\bar{P}_{\rm L}=\bar{N}_{\rm R}=8$   &  $\bar{P}_{\rm R}=\bar{N}_{\rm L}=2$  \bigstrut \\ \hline
\multicolumn{2}{c}{$M=25$ (truncation parameter)}  \bigstrut \\ \hline
\hline
\end{tabular}
\end{center}
\label{tab_values}
\end{table}

\subsection{DMRG Calculations}

\begin{figure}
\centering
\begin{minipage}[t]{0.6\hsize}
\resizebox{1.0\hsize}{!}{\includegraphics{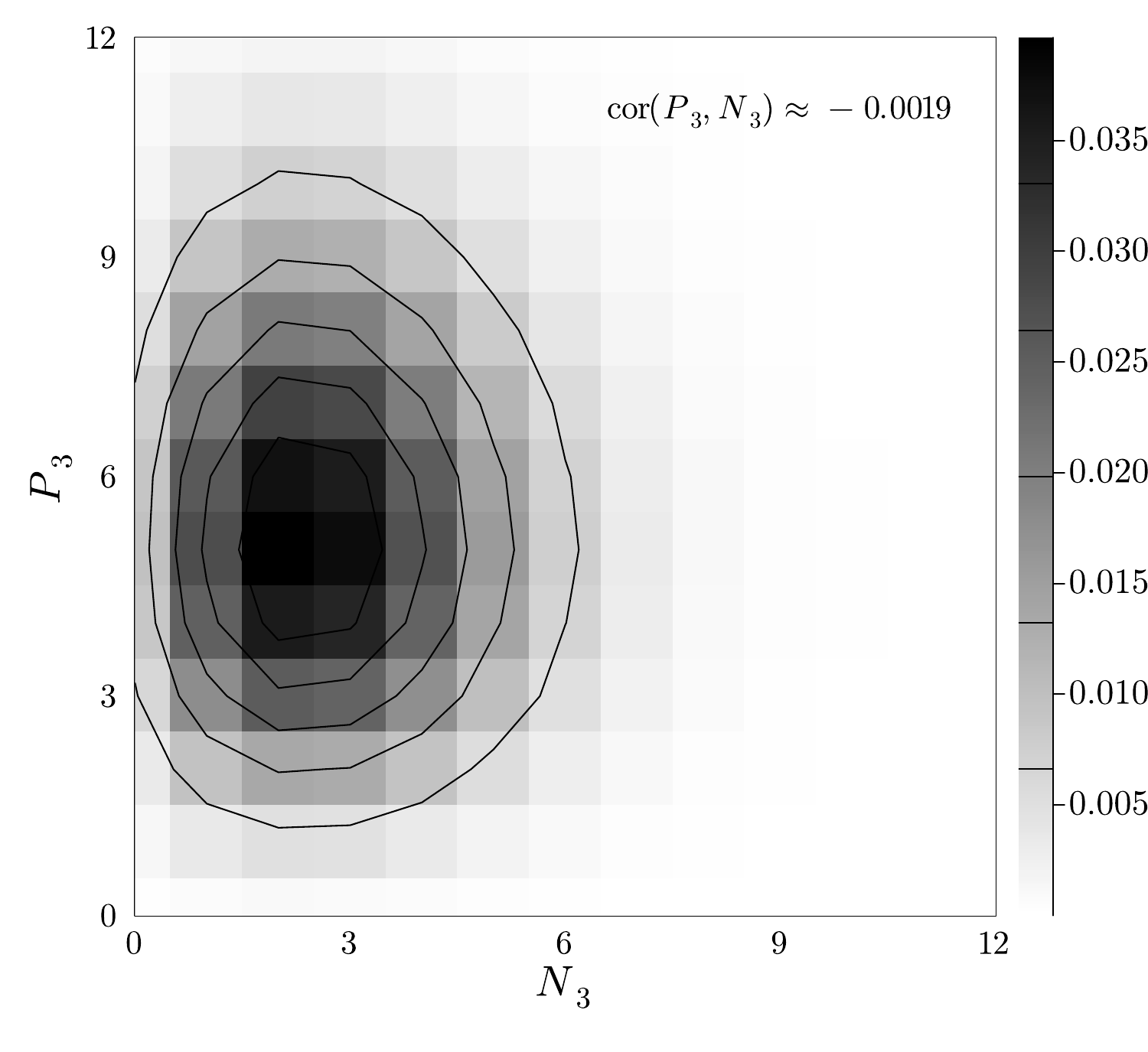}}
\end{minipage}
\caption{Grayscale map of the steady-state probability distribution of holes and electrons $\{P_3,N_3\}$ in the third cell. Five contours and a gray bar indicating the probability values are shown. The parameter values listed in Table~\ref{tab_values} plus $k_+=k_-=10.0$ are used in tensor-network calculations. The annotation is the Pearson correlation coefficient between the probability distributions of $P_3$ and $N_3$.}
\label{fig_P}
\end{figure}

\par To characterize the current fluctuations, we perform full counting statistics of transfers of the unit charge flowing from the left reservoir into the system. We introduce the extended probability distribution ${\cal P}({\bf P},{\bf N},Z,t)$ by including the charge transfers $Z$ in the time interval $[0,\,t]$ and, furthermore, the function $F_{\lambda}({\bf PN},t)\equiv\sum_Z{\cal P}({\bf P},{\bf N},Z,t)\,{\rm e}^{-\lambda Z}$ in terms of the counting parameter $\lambda$. Then, the equation of time evolution,
\begin{align}
\frac{\rm d}{{\rm d}t}F_{\lambda}({\bf PN},t)=\hat{L}_{\lambda}F_{\lambda}({\bf PN},t) \label{eq_evolution}
\end{align}
can be derived~\cite{Lebowitz_JStatPhys_1999}. The solution to this equation is formally given by $F_{\lambda}({\bf PN},t)=\exp(\hat{L}_{\lambda}t)F_{\lambda}({\bf PN},0)\sim\sum_i\exp(l_{\lambda,i}t)$, where $\{l_{\lambda,i}\}$ are the eigenvalues of the tilted generator $\hat{L}_{\lambda}$. In the long-time limit, the term with the leading eigenvalue dominates the solution.  According to its definition~(\ref{eq_Q}), the cumulant generating function for the current statistics is now given by
\begin{align}
Q(\lambda)\equiv\lim_{t\to\infty}-\frac{1}{t}\ln\sum_{\bf PN}F_{\lambda}({\bf PN},t) \text{,}
\end{align}
which is exactly equal to the negative of the leading eigenvalue of the tilted generator $\hat{L}_{\lambda}$.

\begin{figure}
\centering
\begin{minipage}[t]{0.6\hsize}
\resizebox{1.0\hsize}{!}{\includegraphics{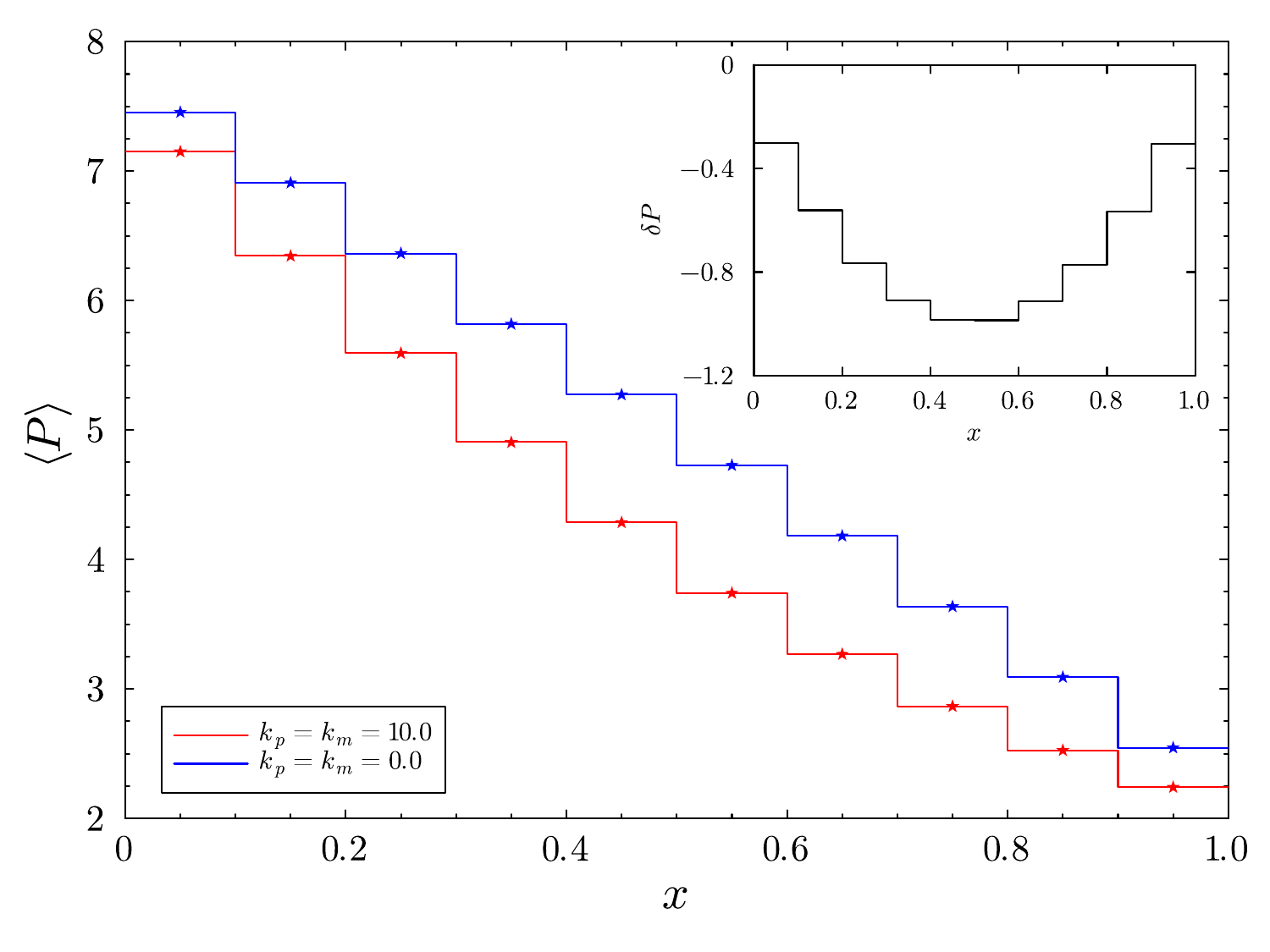}}
\end{minipage}
\caption{The profiles of the average occupation number of holes $\langle P\rangle$ along the system. The lines are results from tensor-network calculations, while the asterisks from solving the ODEs~(\ref{eq_ode1})-(\ref{eq_ode4}). The steplike shape of the lines originates from the discretization in space. The boundary conditions used for solving the ODEs are $p(-0.05)=\bar{p}_{\rm L}$, $n(-0.05)=\bar{n}_{\rm L}$, $p(1.05)=\bar{p}_{\rm R}$, and $n(1.05)=\bar{n}_{\rm R}$. The purpose of shifting the boundaries is to obtain consistent results with those from tensor-network calculations in the discretized scheme. Two cases where the reactions are turned on ($k_+=k_-=10.0$) and off ($k_+=k_-=0$) are shown for comparison. The inset shows the deviation between the two profiles. Other parameter values are given in Table~\ref{tab_values}.}
\label{fig_profile}
\end{figure}

\par Now, we illustrate how to numerically determine the leading eigenvalue of $\hat{L}_{\lambda}$ using the DMRG algorithm. For this purpose, we first represent an arbitrary function $F_{\lambda}({\bf PN})$ as an MPS $\ket{{\bf F}_{\lambda}}$, and represent the tilted generator $\hat{L}_{\lambda}$ as an MPO (see Appendix~\ref{app_MPSMPO} for detailed account). In the MPS representation of $F_{\lambda}({\bf PN})$, there is a total of $L$ sites, with composite physical index $P_iN_i$ for each site, see the left panel in Fig.~\ref{fig_TN}. After truncation, $P_i$ and $N_i$ both take the $M$ possible values $0,\,1,\,\cdots,\,M-1$. Then $P_iN_i$ is mapped to take $M^2$ possible values. Accordingly, the local operators, e.g., $a_i^-$ and $c_i^+$, have a matrix form of dimension $M^2\times M^2$. In the DMRG calculation, we variationally optimize the MPS $\ket{{\bf F}_{\lambda}}$ to find the corresponding the right leading eigenstate $\ket{{\bf\Psi}_{\lambda}}$ such that $\hat{L}_{\lambda}\ket{{\bf\Psi}_{\lambda}}=-Q(\lambda)\ket{{\bf\Psi}_{\lambda}}$. This eigenstate is also normalized according to the rule~(\ref{eq_normalization}), i.e., $\braket{{\bf\Psi}_{\lambda}^{\dagger}|{\bf\Psi}_{\lambda}}=1$. Thus, the cumulant generating function can be calculated as
\begin{align}
Q(\lambda)=-\braket{{\bf\Psi}_{\lambda}^{\dagger}|\hat{L}_{\lambda}|{\bf\Psi}_{\lambda}} \text{,}
\end{align}
where it is worth mentioning that $\bra{\bf\Psi_{\lambda}^{\dagger}}$ does not correspond the left leading eigenstate due to the non-Hermitian of $\hat{L}_{\lambda}$. In addition, the steady state distribution of the number of holes and electrons can be calculated as ${\cal P}_{\rm st}({\bf P},{\bf N})=\Psi_0({\bf PN})/\sum_{\bf PN}\Psi_0({\bf PN})$, which is normalized in probability. Readers are referred to Ref.~\cite{Gu_NewJPhys_2022} for a detailed exposition of the DMRG algorithm applied in a nonequilibrium diffusion system. Some programming details with the ITensors library are provided in Appendix~\ref{app_code}.

\section{Results}\label{sec_results}

\par The numerical results obtained from tensor-network calculations are presented in this section. Some adopted parameter values are listed in Table~\ref{tab_values}. The boundary conditions are indeed symmetric under inversion and permutation between holes and electrons, i.e., $\bar{P}_{\rm L}=\bar{N}_{\rm R}$ and $\bar{P}_{\rm R}=\bar{N}_{\rm L}$, so that there is no ambiguity regarding the affinity. The numbers of holes and electrons are allowed to take possible values from $0$ to $M-1=24$ after truncation, indicating that the number of states in a single cell is $625$. The system of composite $L=10$ cells has a total of $625^{10}$ states. This is an extraordinarily large number that justifies the necessary use of tensor networks. For convenience, all parameter values in this work are presented with no units. In fact, dimensionless quantities can be properly defined; see Appendixes of Refs~\cite{Gu_PhysRevE_2018, Gu_PhysRevE_2025a} for details.

\begin{figure}
\centering
\begin{minipage}[t]{0.6\hsize}
\resizebox{1.0\hsize}{!}{\includegraphics{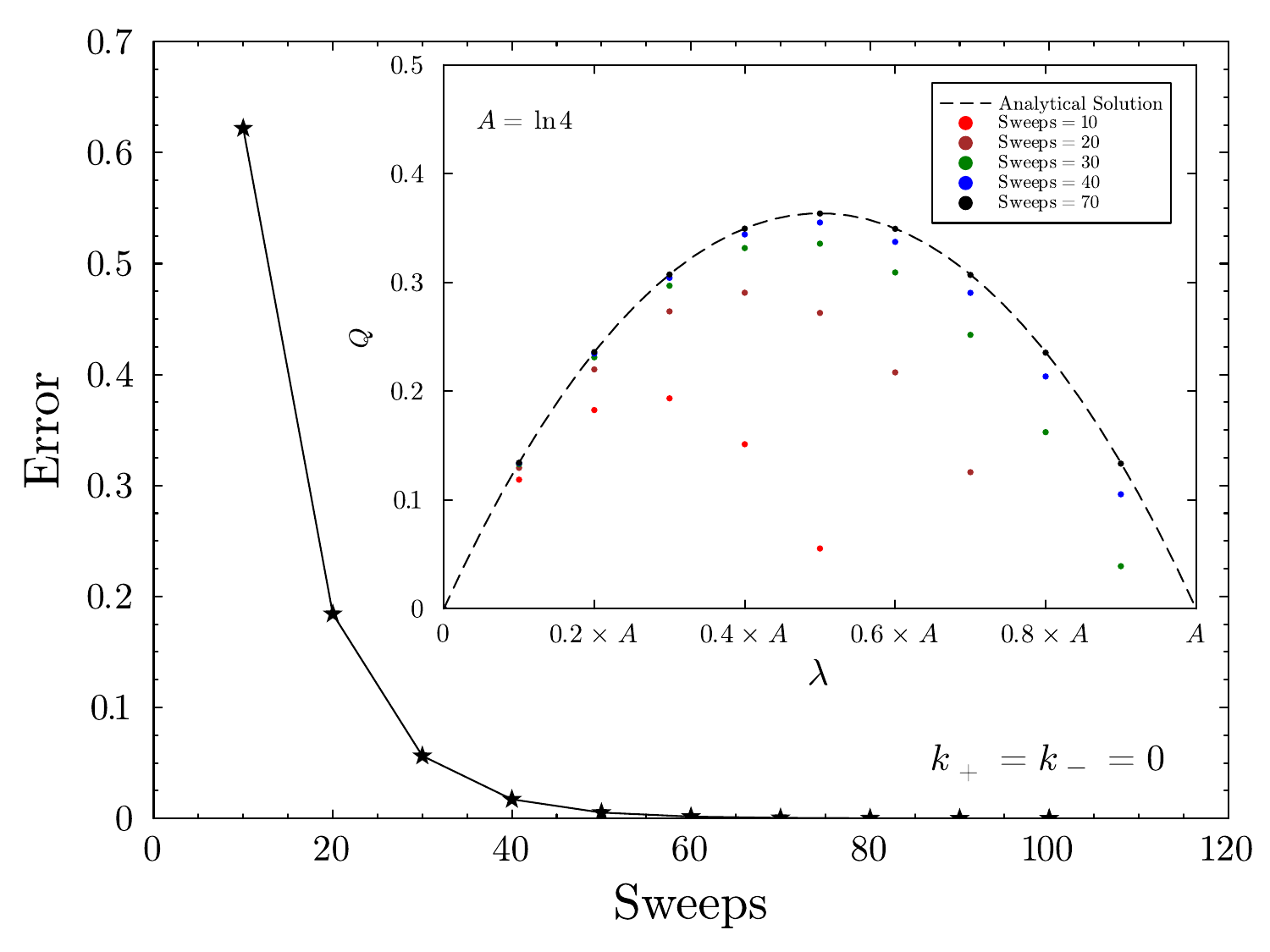}}
\end{minipage}
\caption{Convergence of the cumulant generating function calculated with DMRG as the the number of sweeps is increased. The reactions are turned off, $k_+=k_-=0$, and in this case, analytical solution can be obtained, shown as the dashed line in the inset. The dots represent results calculated using the DMRG algorithm. Different colors correspond to the cases with different numbers of sweeps. The asterisks with solid line joining them represent the errors that are root mean square of the differences between the results calculated using the DMRG algorithm and the analytical solution. The parameter values in Table~\ref{tab_values} are taken. Prior to DMRG calculations, the system is initialized to the steady state such that the number of holes or electrons is Poisson distributed in each cell with the mean value determined from the linear profile shown as the blue line in Fig~\ref{fig_profile}. In DMRG calculations with a high-performance laptop, it took about 380 s with 100 sweeps to compute one data point of $Q(\lambda)$. This time cost may vary depending on the computing platform and the detailed DMRG implementation.}
\label{fig_convergence}
\end{figure}

\par We first calculate the steady state probability distribution of the system ${\cal P}_{\rm st}({\bf P},{\bf N})$. In Fig.~\ref{fig_P}, we show the marginal probability distribution of holes and electrons in the third cell in grayscale map. Five contours are also marked. In this figure, there is no apparent statistical correlation between holes and electrons, as also confirmed quantitatively by the negligible Pearson correlation coefficient shown in the figure. However, we point out that there should be potentially positive correlation since pairs of holes and electrons are generated and recombined stochastically through the reaction process. The absence of this is explained by the transition of excess holes or electrons to neighboring cells. In this figure, we also observe that the probabilities of ${\cal P}(P_3,N_3)$ become vanishingly small when $P_3$ and $N_3$ tend to $M-1=24$, justifying the choice of truncation parameter $M=25$ in Table~\ref{tab_values}. We then calculate the average occupation number $\langle P_i\rangle=\sum_{\bf PN}P_i{\cal P}({\bf PN})$ of holes along the system, as shown in Fig.~\ref{fig_profile}. In this figure, we compare two cases where the reactions (controlled by rate constants $k_+$ and $k_-$) are turned on or off. When the reactions are turned off, the holes undergo the process of pure diffusion. In this case, the profile should be linear interpolating the hole concentrations in two boundary reservoirs. This is indeed observed. However, when the reactions are turned on, the profile is bent, as shown in the figure. The profile $p(x)$ can also be obtained by solving the ODEs~(\ref{eq_ode1})-(\ref{eq_ode4}). The discretized samples are also displayed in this figure, and we see that they are in agreement with the results from tensor-network calculations, sufficiently supporting the validity of tensor networks applied in the diffusion-reaction systems.

\begin{figure}
\centering
\begin{minipage}[t]{0.6\hsize}
\resizebox{1.0\hsize}{!}{\includegraphics{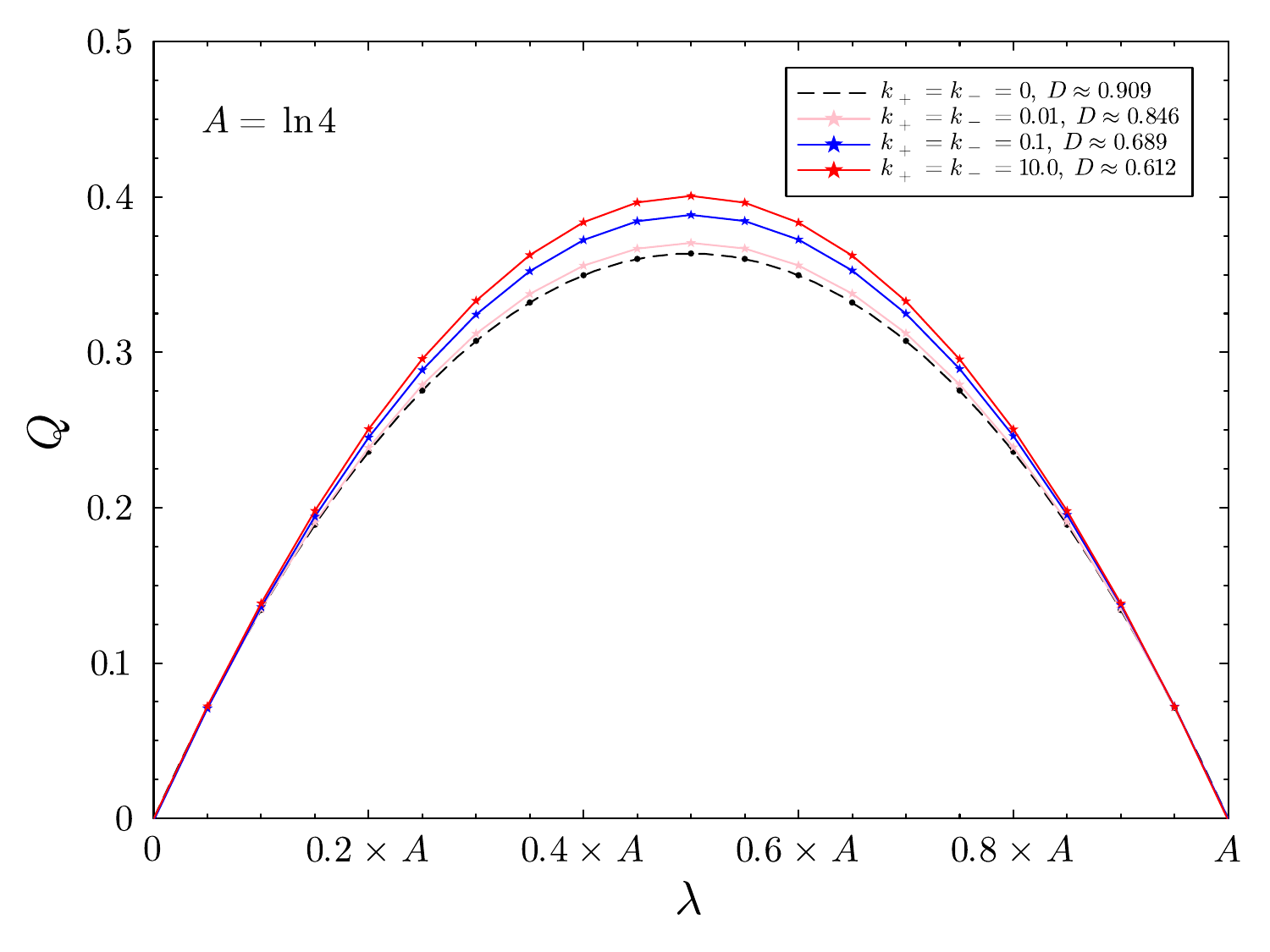}}
\end{minipage}
\caption{Cumulant generating function. The asterisks with solid line joining them are results for the system parameter values listed in Table~\ref{tab_values} and, respectively, $k_+=k_-=0.01,\,0.1,\,10.0$. The dots are results with values listed in Table~\ref{tab_values} and $k_+=k_-=0$. The dashed line is plotted from the analytical solution~(\ref{eq_Q_solution}) with the same parameter values as those for dots. $A=\ln 4$ is the affinity calculated from Eq.~(\ref{eq_A}). In DMRG calculations, $100$ sweeps are taken to guarantee convergence of the results. The legend also shows the values of the diffusivity calculated with numerical differentiations from $Q(\lambda)$ according to Eq.~(\ref{eq_D}).}
\label{fig_Q}
\end{figure}

\par The main result of this work is the tensor-network calculation of the cumulant generating function $Q(\lambda)$, which is presented in both Fig.~\ref{fig_convergence} and Fig.~\ref{fig_Q}. In DMRG calculations, the number of sweeps is crucial for cumulant generating function to converge. When the reactions are turned off ($k_+=k_-=0$), the charge transport is reduced to the pure diffusion of holes and electrons. In this case, $Q(\lambda)$ can be analytically solved~\cite{Andrieux_JStatMech_2006, Gu_NewJPhys_2022} with the solution given by
\begin{align}
Q(\lambda)= & \frac{k\bar{P}_{\rm L}}{L+1}\left(1-{\rm e}^{-\lambda}\right)+\frac{k\bar{P}_{\rm R}}{L+1}\left(1-{\rm e}^{+\lambda}\right) + \frac{k\bar{N}_{\rm L}}{L+1}\left(1-{\rm e}^{+\lambda}\right)+\frac{k\bar{N}_{\rm R}}{L+1}\left(1-{\rm e}^{-\lambda}\right) \text{.} \label{eq_Q_solution}
\end{align}
Moreover, in this purely diffusive case, the mean current and its diffusivity can be evaluated from Eq.~(\ref{eq_Q_solution}) according to Eqs.~(\ref{eq_J}) and (\ref{eq_D}),
\begin{align}
& J=\frac{k}{L+1}\left(\bar{P}_{\rm L}-\bar{P}_{\rm R}+\bar{N}_{\rm R}-\bar{N}_{\rm L}\right) \text{,} \\
& D=\frac{k}{2(L+1)}\left(\bar{P}_{\rm L}+\bar{P}_{\rm R}+\bar{N}_{\rm L}+\bar{N}_{\rm R}\right) \text{.} \label{eq_DD}
\end{align}
Recall that the affinity $A$ is given by Eq.~(\ref{eq_A}); we can immediately obtain the equality
\begin{align} 
\frac{2D}{J}=\coth(A/2) \text{,} \label{eq_equality}
\end{align}
which holds in this purely diffusive case. In Fig.~\ref{fig_convergence}, the results calculated using the DMRG algorithm in the purely diffusive case are presented. The parameter values are listed in Table~\ref{tab_values}. In this figure, the results calculated with different numbers of sweeps are shown for comparison. We notice that $70$ sweeps are sufficient to guarantee the convergence to the analytical solution. The error vanishes quickly with the number of sweeps. Thus, the results can be considered numerically exact if more than $70$ sweeps are taken in DMRG calculations. This shows the advantage of tensor-network methods over the Monte Carlo simulation methods for calculating the large deviations in many-body systems. With the latter class of methods, such as cloning approach, the cumulant generating function of the current statistics in the diffusion-reaction system cannot be easily computed accurately, especially in the region around $\lambda=A$. Moreover, excellent agreement between numerical results and the analytical solution manifests the validity of applying tensor networks to this kind of nonequilibrium system. In Fig.~\ref{fig_Q}, $Q(\lambda)$ is plotted in several cases: $k_+=k_-=0,\,0.01,\,0.1,\,10.0$, all with the same other parameter values listed in Table~\ref{tab_values}. It is obvious that $Q(\lambda)$ satisfies the Gallavotti–Cohen symmetry~(\ref{eq_FT2}) in all these cases, as expected. When $k_+=k_-=0.01,\,0.1,\,10.0$, we do not have analytical solutions, and in these cases we can only rely on the numerical values of $Q(\lambda)$ to analyze the current fluctuations. From Fig.~\ref{fig_Q}, it seems that the $Q(\lambda)$ have the same slope at $\lambda=0$ for all cases. This directly indicates that the system has the same mean current $J$ whether or not there are reactions between holes and electrons. With a bit more analysis, it is indeed so. Suppose that the holes and electrons have the linear profiles $p(x)$ and $n(x)$ in the case $k_+=k_-=0$. When the reactions are turned on, the profiles are deformed, becoming $p(x)+\delta p(x)$ and $n(x)+\delta n(x)$, respectively. Because the reactions always lead to generation or recombination of holes and electrons in pairs and these charge carriers are also assumed to have the same diffusion coefficients, it is arguably that $\delta p(x)=\delta n(x)$. As depicted in the inset of Fig.~\ref{fig_profile}, the deviation between the two hole profiles corresponding to the cases where the reactions are turned on and off are found symmetric. Moreover, the profiles for holes and electrons should also be symmetric due the proper boundary conditions. This suggests that the same deviation $\delta p(x)=\delta n(x)$ is indeed true, supported by numerical data. The mean current is calculated as follows:
\begin{align}
J & =\langle j_p(x)-j_n(x)\rangle \nonumber \\
& =-D_p\partial_x[p(x)+\delta p(x)]+D_n\partial_x[n(x)+\delta n(x)] \nonumber \\
& =-D_p\partial_xp(x)+D_n\partial_xn(x) \text{,} \label{eq_J2}
\end{align}
where the third line is obtained due to the assumption $D_p=D_n$ together with $\delta p(x)=\delta n(x)$. This indicates that the mean current indeed remains the same for all cases. It should be noted that Eq.~(\ref{eq_J2}) holds due to the symmetry of the boundary conditions. In Fig.~\ref{fig_Q}, we also compare the trends of slope change (second-order derivatives) of the $Q(\lambda)$ at $\lambda=0$ for all cases. It is apparently discernible that the current diffusivity given by Eq.~(\ref{eq_D}) is smaller in the cases $k_+=k_-=0.01,\,0.1,\,10.0$ than that in the case $k_+=k_-=0$. The values of the diffusivity obtained from numerical differentations in the legend also confirm this observation. It is particularly noteworthy that $D\approx 0.909$ in the case of $k_+=k_-=0$ is in agreement with the value calculated from Eq.~(\ref{eq_DD}). Thus, starting from the equality~(\ref{eq_equality}), we arrive at an interesting inequality
\begin{align} 
\frac{2D}{J}\le\coth(A/2) \text{,} \label{eq_inequality}
\end{align}
where the equality sign holds when there is no reaction or equivalently when the currents of holes and electrons are not coupled. In this case, the current arises from pure diffusion directed by imbalanced boundary conditions. Since the system evolves according to the master equation~(\ref{eq_master_equation}), we can simulate the system with stochastic methods. In this way, we can also evaluate the diffusivity of the current. However, we note that the stochastic simulation is time consuming and the result is intrinsically prone to statistical error. In comparison, the current diffusivity evaluated from the cumulant generating function shown in Fig.~\ref{fig_Q} is more convincing. The essential reason lies in that the constraint by the Gallavotti-Cohen symmetry~(\ref{eq_FT2}) is taken for advantage of to show the trend of slope change at the origin. For the 1D diffusion-reaction system under study, it should be emphasized again that the inequality~(\ref{eq_inequality}) is established only when the affinity can be consistently defined, i.e., the symmetric boundary conditions~(\ref{eq_constraints}) are imposed. In this work, we have assumed for simplicity the equality between the diffusion coefficients of holes and electrons, $D_p=D_n$. This simplifying condition together with symmetric boundary conditions results in the unaltered mean current whether or not the reaction is turned on or off. This allows us to focus solely on the current fluctuations whose value can be derived from the second-order derivatives of $Q(\lambda)$ at $\lambda=0$. Since the mean current is unchanged, the comparison of the current fluctuations between different cases of reaction can be easily made; see Fig.~\ref{fig_Q}. Here, we point out that the inequality~(\ref{eq_inequality}) remains valid even if the condition $D_p=D_n$ is discarded. There are two reasons for this: (1) When the reaction is turned off, the equality~(\ref{eq_equality}) can still be established, only if the system has symmetric boundary conditions, and (2) when the reaction is turned on, the consequence due to the reaction is the same whether or not both diffusion coefficients are equal to each other.

As to why current fluctuations are suppressed by the reactions, we now provide a heuristic explanation. According to Eqs.~(\ref{eq_ode3}) and (\ref{eq_ode4}), the balance between hole-electron pair generation and recombination requires that $k_+=k_-p(x)n(x)$. This condition is satisfied at the boundaries with the parameter values in Table~\ref{tab_values} together with $k_+=k_-$, which are set in numerical simulations. We now consider the case where the reactions are turned on. If holes and electrons had linear profiles in this case, then their densities in the middle of the system would be $p(l/2)=n(l/2)=0.5\times(8+2)/4=1.25$ according to the parameter values in Table~\ref{tab_values}. This directly leads to more hole-electron pairs being recombined than generated, $k_-p(l/2)n(l/2)>k_+$. As such, the true profiles are bent that the charge densities become smaller, as indeed observed in Fig.~\ref{fig_profile}. Moreover, we gain insight from Eq.~(\ref{eq_DD}) that the current diffusivity is intuitively determined by the sum of charge carriers, not the difference. Combining this insight with the true profile in Fig.~\ref{fig_profile} heuristically explains the suppressed current fluctuations. The key ingredient is the nonlinear reactions that tend to deform the linear profiles dictated by the diffusion processes. This nonlinearity in rates is ubiquitous in chemical reactions according to the law of mass action. However, it remains an open question what exactly the conditions on the reaction terms are such that they lead to reduced fluctuations.

The inequality~(\ref{eq_inequality}) indicates that the current fluctuations are upper bounded, complementing the thermodynamic uncertainty relation (TUR)~\cite{Barato_PhysRevLett_2015, Gingrich_PhysRevLett_2016}, which states that the current fluctuations are lower bounded. When the system is close to equilibrium, the inequality~(\ref{eq_inequality}) should reduce to the fluctuation-dissipation relation $J=DA$, implying that the upper bound dictated by the inequality~(\ref{eq_inequality}) is asymptotically attained in the linear regime. Moreover, the TUR first reported in Ref.~\cite{Barato_PhysRevLett_2015} can be reformulated in the present context as $D\ge J/A$, where the lower bound tends to coincide with the upper bound given by the inequality~(\ref{eq_inequality}) in the limit $A\to 0$. This consistently gives the fluctuation-dissipation relation $J=DA$. The inequality~(\ref{eq_inequality}) has also been reported by us previously in systems of quantum ballistic transport, the symmetric simple exclusion process, and the transport process of charged particles driven by electrostatic fields~\cite{Gu_PhysRevE_2025c}. In particular, the inequality~(\ref{eq_inequality}) was exactly proven for a broad class of systems of quantum ballistic transport. In this work, the long-range electrostatic interactions between charge carriers are neglected. This is for the convenience of using tensor networks to study this diffusion-reaction system. However, in Ref.~\cite{Gu_PhysRevE_2025c}, the charge transport driven by the electrostatic fields shows that the inequality~(\ref{eq_inequality}) remains valid. Thus, we can be assured that in the diffusion-reaction system under study the inequality~(\ref{eq_inequality}) still holds even if the electrostatic interactions are taken into account. For previously studied systems, the repulsive interactions between transport particles were interpreted as dampening the current fluctuations. In the diffusion-reaction system studied in this work, the reactions have the same effect on current fluctuations.

\section{Conclusion and Perspectives}\label{sec_conclusion}

We have presented in detail how to compute the cumulant generating function of the current statistics in a 1D nonequilibrium diffusion-reaction system with tensor networks. By comparing the cases where the reactions are turned on and off, we demonstrate that the reactions between holes and electrons dampen the current fluctuations, indicating an interesting inequality that could be of fundamental importance. We hope this may raise interest among peer scientists and lead to more investigations in this regard. One possible exploration would be extending the inequality in driven diffusive systems where the affinity is not well defined, e.g., in the diffusion-reaction system with no symmetric boundary conditions. Besides, our work adds to the continuously expanding applications of tensor networks to study the dynamical fluctuations in classical stochastic systems, providing significant promise for the tensor networks to become standard tools in this field of nonequilibrium statistical physics. In prospect, tensor networks may be extended to study the diffusion-reaction systems in two dimensions to explore the the behavior of dynamical patterns. In addition, since the transition rates in the master equation for the system under study all depend on the local number of holes and electrons, future explorations may also consider systems where transition rates are jointly determined by the states of a few of neighboring sites.

\section*{Acknowledgments}
The author thanks P. Gaspard for his helpful remarks on the manuscript. This work was supported by the National Natural Science Foundation of China (NSFC) under the Grant No. 12505048.

\appendix

\section{Matrix Product State and Matrix Product Operator}\label{app_MPSMPO}

The steady state determined by the equation of time evolution~(\ref{eq_evolution}) can be effectively represented by a matrix product state. This ansatz can be justified by considering the case where the holes and electrons are decoupled, i.e., the reaction is turned off. In this scenario, the steady state is given by~\cite{Andrieux_JStatMech_2006}
\begin{align}
F_{\lambda,{\rm st}}({\bf PN},t)={\rm e}^{-Q(\lambda)t}\prod_{i=1}^L\frac{A_i^{P_i}}{P_i!}\cdot\frac{B_i^{N_i}}{N_i!} \text{,}
\end{align}
where $\{A_i\}$ and $\{B_i\}$ are $\lambda$-dependent constants. When $\lambda=0$, the cumulant generating function is zero, $Q(0)=0$, and the constants are the mean numbers of holes or electrons, $A_i=\langle P_i\rangle$ and $B_i=\langle N_i\rangle$, which can be determined by linearly interpolating the particle concentrations between two reservoirs. The steady state when $\lambda=0$ is nothing but the product of Poisson distributions, differing only by a renormalization constant. There is no correlation between holes or electrons in adjacent cells. Thus, the representation of matrix product state is the one such that the bond dimension is simply one. When the reaction is turned on, we believe that a matrix product state with small bond dimension suffices to represent the steady state. In practical DMRG calculations with ITensors library, both parameter values \texttt{maxdim=100} and \texttt{cutoff=1.0e-18} are adopted. The former one means that the maximum dimension takes $100$ when singular value decompositions are performed. The latter denotes the condition $\sum_{n\in{\rm discarded}}\sigma_n^2/\sum_n\sigma_n^2<1.0^{-18}$, where $\{\sigma_n\}$ are singular values. These parameter values are sufficient to guarantee the accuracy.

The transition rates in the master equation~(\ref{eq_master_equation}) can be expressed in terms of two local operators in the Doi-Peliti formalism. This allows us to write the tilted generator in the form~(\ref{eq_tilted_generator}), which can be represented as a large matrix in principle. However, we notice that in Eq.~(\ref{eq_tilted_generator}) the tensor product operations are between adjacent local operators. This nearest-neighbor property makes it possible for the tilted generator to be represented compactly as an MPO, which can be seen as a generalized version of a matrix. This MPO representation is the most crucial step in implementing the DMRG algorithm. In the following, we explicitly give the tilted generator in the form of an MPO, reading
\begin{equation}
\resizebox{0.93\hsize}{!}{$
\hat{L}_{\lambda}=\begin{pmatrix}
O_0 & a_0^-{\rm e}^{-\lambda} & a_0^+ & b_0^-{\rm e}^{+\lambda} & b_0^+ & 1
\end{pmatrix}\otimes\begin{pmatrix}
1 & 0 & 0 & 0 & 0 & 0 \\
a_1^+ & 0 & 0 & 0 & 0 & 0 \\
a_1^-{\rm e}^{+\lambda} & 0 & 0 & 0 & 0 & 0 \\
b_1^+ & 0 & 0 & 0 & 0 & 0 \\
b_1^-{\rm e}^{-\lambda} & 0 & 0 & 0 & 0 & 0 \\
O_1 & a_1^- & a_1^+ & b_1^- & b_1^+ & 1
\end{pmatrix}\otimes\cdots\otimes\begin{pmatrix}
1 & 0 & 0 & 0 & 0 & 0 \\
a_i^+ & 0 & 0 & 0 & 0 & 0 \\
a_i^- & 0 & 0 & 0 & 0 & 0 \\
b_i^+ & 0 & 0 & 0 & 0 & 0 \\
b_i^- & 0 & 0 & 0 & 0 & 0 \\
O_i & a_i^- & a_i^+ & b_i^- & b_i^+ & 1
\end{pmatrix}\otimes\cdots\otimes\begin{pmatrix}
1 \\
a_{L+1}^+ \\
a_{L+1}^- \\
b_{L+1}^+ \\
b_{L+1}^- \\
O_{L+1}
\end{pmatrix}$} \text{,} \label{eq_MPO}
\end{equation}
where
\begin{align}
O_i=\begin{cases}
-a_0-b_0 & \text{for}\hspace{0.5cm} i=0 \text{,} \\
-2a_i-2b_i+c_i^++c_i^--c_i & \text{for}\hspace{0.5cm} i=1,\cdots,L \text{,} \\
-a_{L+1}-b_{L+1} & \text{for}\hspace{0.5cm} i=L+1 \text{.}
\end{cases}
\end{align}
There is a total of $L+2$ tensors in the MPO~(\ref{eq_MPO}), with the first and last tensors acting on the two reservoir cells, respectively. According to Eqs.~(\ref{eq_bc1})-(\ref{eq_bc5}), the individual operators composing these two tensors, such as $a_0^-$, $a_0$, and $b_{L+1}$, are all constants (order-0 tensors). They can be contracted with their the neighboring tensors, leading to a resultant MPO of length $L$, in agreement with that of MPS. The MPO representation of the tilted generator in Eq.~(\ref{eq_MPO}) is most compact, with a minimal bond dimension of $6$. This compactness is as the same as the MPO representation automated by the ITensors library~\cite{Fishman_SciPostPhysCodeb_2022}.

\section{Julia Programming with ITensors Library}\label{app_code}

\par \textsc{julia} is a high-level and a dynamic programming language that makes it user interactive~\cite{Bezanson_SIAMRev_2017}. It is on the track to becoming one of the most welcomed languages in scientific high performance computing. ITensors is a \textsc{julia} library for rapidly creating correct and efficient tensor network algorithms~\cite{Fishman_SciPostPhysCodeb_2022}. It allows users to focus the connectivity of tensor networks without manually recording and tracking tensor indices. The indices of a tensor have unique identities. When contracting two tensors, matching indices find each other and contract automatically. The ITensors library was developed mainly for quantum many-body systems. It comes with several built-in site types for quantum systems e.g., \verb"S=1/2" and \verb"Fermion." However, for the classical system considered in this work, new site type should be manually defined. In the following, a brief guide is provided on how to achieve this. The code line \texttt{ITensors.space(::ITensors.SiteType"DR")=M*M} defines a site type called \texttt{DR} (Diffusion-Reaction) with the dimension \texttt{M*M}, where \texttt{M} is the local truncation parameter in Table~\ref{tab_values}. Then we can define the site lattice of type \texttt{DR} and length \texttt{L} with the code line \texttt{sites=ITensors.siteinds("DR", L)}. The local operators can be defined for this site type. For example, the following code lines define the identity operator with the name \texttt{I} and dimension $M^2\times M^2$:
\begin{verbatim}
I=Base.zeros(Float64, M*M, M*M)
for i in 0:M-1
        for j in 0:M-1
                I[i*M+j+1, i*M+j+1]=1.0
        end
end
ITensors.op(::ITensors.OpName"I", ::ITensors.SiteType"DR")=I
\end{verbatim}
where \texttt{i} denotes the hole number and \texttt{j} denotes the electron number. Note that array indices in \textsc{julia} start from $1$. Other local operators can be defined in a similar way. With all local operators available, the tilted generator can be given by \texttt{opsum=ITensors.OpSum()} and \texttt{opsum+=()} in a manner similar to quantum Hamiltonians. The MPO representation of the tilted generator can also be constructed by \texttt{mpo=ITensors.MPO(opsum, sites)}. The built-in DMRG routine is not applicable for this customized system. Instead, one should manually write code for it. A detailed account can be found in Ref.~\cite{Gu_NewJPhys_2022}. The code for this work is available upon reasonable request.

\printbibliography[title={References}]

\end{document}